\documentstyle[preprint,epsfig,aps,psfig]{revtex}
\begin{document}
\tightenlines
\title{Phase behaviour of a hard sphere colloidal system in the presence
of an external laser field}  
\author{Pinaki Chaudhuri, Ajay K. Sood, H. R. Krishnamurthy} 
\address{ Department of Physics\\ Indian Institute of Science\\
Bangalore 560012, India.\\}
\maketitle
\narrowtext
\begin{abstract}

We study the phase behaviour of a two-dimensionally confined hard sphere colloidal system
in the presence of a periodic light field of two interfering laser beams
using Monte Carlo simulations. For a given packing fraction of 
the particles,
the colloidal system undergoes a transition from a modulated liquid to a modulated
crystal as the light intensity is increased, corresponding to laser 
induced
freezing. For certain packing fractions of the colloidal particles, the system 
again
becomes a modulated liquid as the field strength crosses a threshold value,
showing a re-entrant behaviour.

\end{abstract}

\narrowtext
\newpage
In recent times, there has been a renewed interest in laser induced
freezing (LIF), i.e, the freezing of 
colloidal particles in the presence of an external
modulating (laser) field. LIF 
 was first demonstrated by Choudhury {\it et al}
 \cite{clark}. A sample of 1$\mu$m diameter polysterene 
particles was suspended in water placed between two optical flats. 
The highly charged particles, repelled by the walls, formed a two-dimensional monolayer structure. This two-dimensional system was subjected to a laser
intensity pattern periodically modulated along one direction. It was found that 
when the wavevector of the stationary electric field is tuned to half the
wave vector $q_{o}$ where the liquid structure factor peaks, a triangular lattice with full two-dimensional symmetry results for light intensities above a threshold value. 

A simple theoretical analysis in terms of the Landau-Alexander-McTague \cite{clark,lmt} theory showed a rich phase diagram with a first order
freezing transition turning to second order at higher field strenghts via a tricritical point. In a density functional theory \cite{ry} analysis Chakrabarti
{\it et al} \cite{hrk} obtained a similar cross over and in addition deduced
 a symmetry
criteria for obtaining the tricritical point in the phase diagram.
Monte Carlo simulational studies of $1.1 \mu m$ diameter colloidal particles
interacting via a purely repulsive DLVO potential by Chakrabarti {\it et al}
\cite{jay} semed to confirm such a phase diagram, showing in addition an interesting 
reentrant transition to a liquid state for higher laser fields.
 Recently Wei {\em et al} \cite{wei} reported an experimental study of LIF 
and from the real space density plots concluded that there is indeed
such a re-entrant liquid phase. However, more recent Monte-Carlo
simulations by Das {\em et al } \cite{cdas1,cdas2} of the same system
as in \cite{jay} did not show any re-entrant phase.
An even more recent theoretical work by Frey {\it et al} \cite{frey},  argues
(by extending the KTHNY picture of unbinding of dislocation pairs during
melting in two dimensions to the case when an external field
is present) that for
highly-screened short-range interactions, there should exist a re-entrant
transition to a liquid-state. 

In this paper, we study the phase behaviour of a system of hard sphere
colloidal particles confined to a 2-d layer under the 
influence of a modulated laser field as a function of the light intensity.
Using Monte Carlo simulations, we see that as the field intensity is increased, the
system changes from a modulated liquid to a crystal, and finally back to 
a modulated liquid again, i.e, LIF in a hard-sphere colloid system does seem to
show
reentrant laser induced melting transition (LIM) unlike the $1.1 \mu m$ DLVO
colloidal system \cite{cdas1,cdas2}.

Hard sphere systems 
are interesting model systems in their own right \cite{luben}
and should be a reasonable zeroth approximation for describing
charge stabilised colloids which are highly screened.
In such systems, where there is an exclusion of the 
interpenetration of the particles, the inter-particle potential is 
$U(r) = \infty 
 (r < R )$ and
 $U(r) = 0
 (r > R )$
 , where R is the hard sphere radius. 
The free energy is determined entirely by entropy, which is dependent
on the fraction $\phi$ of the total volume occupied by the spheres.
At low volume fractions, the inter-particle collisions are rare. However
when the packing fraction $\phi$ increases, the motion of the particles is
restricted because of more and more collisions with neighbouring particles.
As $\phi$ is increased from the liquid phase, it becomes entropically favourable
for the system to form a periodic crystal rather than a random fluid 
structure. Thus, there is a purely entropy driven liquid-to-solid transition which is interesting \cite{luben}.
The objective of this paper is to study 
the effect of the external field on this entropy driven phase transition.

Our simulations have been done in an ensemble where both the rectangular box size and the number of particles are fixed. In order to change the
the area fraction occupied by the hard spheres, the diameter of the colloidal particles is changed.
Temperature is not a relevant quantity when the external field is absent; but in the
presence of the field, the unit of energy chosen is the thermal energy
$k_{B}T$. We consider the external laser field to be modulated along the
y-direction and model its effects in terms of a potential of the form $ V({\bf
r}) = V_{o}cos(q_{o}y)$,  where $q_{o}$ ,
equal to $2\pi/\frac{a\sqrt3}{2}$, is the smallest reciprocal lattice
vector corresponding to the triangular lattice of lattice
contant $a$. 

The
particles are contained in a rectangular box where the ratio of the two sides
$L_{y}$ and $L_{x}$ is $\sqrt 3 : 2$. This ensures that the simulation box is
commensurate with a triangular lattice. $L_{x}(=L)$ is chosen as the square
root of the number of particles. To avoid surface effects, we have chosen
periodic boundary conditions using minimum image convention. The rectangular
simulation box is replicated throughout space and in the course of the
simulation, as a particle moves in the original box, so do the periodic images
in exactly the same manner. In our similations, we have considered a system
of 900 particles. The typical number of configurations over which
equilibration has been done is of the order of $5\times 10^{6}$.

The thermodynamic quantity of our interest is the translational order parameter.
The translational order parameters are defined by :\\
\begin{equation}
 \rho_{\vec {\bf q}_{k}} = < \frac{1}{N} \sum_{i} exp(i\vec
{\bf q}_{k}. \vec {\bf r}_{i})>
 \end{equation}
In the above equation, $\vec {\bf r}_{i}$ is the real-space co-ordinate of the
{\it i}-th particle, $\vec {\bf q}_{k}$ is the {\it k}-th reciprocal 
lattice vector (rlv) belonging to the first shell of the rlvs. 
The averaging has been done over a canonical ensemble. \\
 The value of $
\rho_{\vec {\bf q}_{k}}$ as defined in eqn (1) depends on the origin of co-ordinates.
Therefore, in our simulations, we measure the translational order parameter as
:\\ 
\begin{equation}
 \rho_{\vec {\bf q}_{k}} = < \frac{1}{N}| \sum_{i}
exp(i\vec {\bf q}_{k}. \vec {\bf r}_{i} )|> 
\end{equation} 
The order parameter so
defined is of order unity in the crystalline state, while in the liquid state it
goes to zero as $\frac{1}{\sqrt N}$ for a system of N particles.\\
 The
primitive translation vectors of the direct triangular lattice can be chosen
as : ${\bf a}_{1} = a(1,0), {\bf a}_{2} =a
(\frac{-1}{2},\frac{\sqrt3}{2})$ and those for the
reciprocal triangular lattice as ${\bf q}_{1} =
\frac{2\pi}{a\sqrt3/2}(\frac{\sqrt3}{2},\frac{1}{2}), {\bf q}_{2} =
\frac{2\pi}{a\sqrt3/2}(0,1)$. In our simulations, we have chosen $a$
to be unity. We denote by
${\bf \rho}_{l}$  the order-parameters for the wavevectors $\pm  {\bf q}_{2}$ 
 parallel to the modulation
wave-vector and ${\bf \rho}_{d}$ denote the order parameter for the other
four first-shell wave-vectors.

In
Fig.\ref{z1}(a) we have plotted ${\bf
\rho}_{d}$ for the cases when there is no external field and 
when the field strength is infinite. 
In the case of zero field, 
entropy is the only mechanism that is forcing the density modes at the
wave vectors $\vec {\bf q}_{k}$ to develop, and this happens at
a packing fraction of about 0.7.
For the case of infinte strength, the 
external potential is so large that the freedom of the particles to
move in a y-direction is totally supressed and hence they can
execute motion only in the x-direction. 
Now we find that ${\bf \rho}_{d}$ becomes
sizeable when the area fraction exceeds 0.68. This value of the area-fraction
corresponds to the situation when there is vertical contact between the particles 
for the first time. For all area fractions less than 0.69, the system consists
of $L$ uncoupled 1-dimensional chains of particles, with particles in the same
line interacting through the hard-sphere repulsive potential. We know that a
disorder-to-order transition is not possible in a 1-dimensional system
with short-ranged interactions. Hence
there is no ordering before $\phi = 0.69$. Development of the density mode corresponding to the wave-vector  ${\bf q}_{2}$ occurs only when a coupling between two
neighbouring chains is established through vertical contact.
To identify the threshold for the transition from the modulated liquid
to modulated crystal phase, we have plotted $d\rho_{d}/d\phi$ as
a function of $\phi$ in Fig. \ref{z1}(b). It can be seen that 
$d\rho_{d}/d\phi$ has a pronounced peak at certain $\phi$
signifying sharp change in $\rho_{d}$ at that value of $\phi$. We 
identify the peak in  $d\rho_{d}/d\phi$  as the onset of 
the freezing transition leading to a non-zero
density mode
$\rho_{d}$. These area fractions for the zero-field and infinite-field are
0.702 and 0.688, respectively. 

In Fig. \ref{t2}, we have plotted the
variation of  ${\bf \rho}_{d}$ and $d\rho_{d}/d\phi$
 with changing area fractions for some typical values of 
field strength ($\beta V_{o} = 0.1, 1.0, 10.0, 2000.0$,
where $\beta=1/k_{B}T$). 
From the plots we have identified, for each value of $\beta V_{o}$,  the corresponding value of
$\phi$ where the onset of freezing occurs. Using these values, we have thus
constructed the phase diagaram of the hard-sphere colloidal system in the
presence of an external laser field, which is plotted in
Fig \ref{phase}.

As noted above, in the
absence of the laser field, the onset of freezing takes place at an area
fraction value of 0.702 . When the field is
switched on, the density modes ($\rho_{l}$) corresponding to the modulating
wave-vector
become nonzero, and this facilitates the deveopment of the 
the other density modes ($\rho_{d}$) causing the freezing transition 
to happen at a smaller $\phi$.
As we can see from the phase
diagram, the value of $\phi$ required for freezing continues to decrease till a $\beta V_{o}$ value of 2.0, where it is down to 0.667. 
Then it increases again to reach the infinite-field limit of about 0.69.
A possible explanation for this latter increase could be the following. As $\beta V_{o}$ gets increased
to values much larger than unity, the motion of the
 particles become more and more restricted to be around
the lines $q_{o}y = (2n+1)\pi$.
Thus, beyond a threshold value
 of the field strength,
 the influence of particles of neighbouring lines on the motion of particles
on a particular line decreases.
This manifests in the fact that the condensation of the density mode  $\rho_{d}$
takes place at a higher area fraction compared to the values corresponding
to field strengths below that threshold. With increasing field strength, the
value of the critical $\phi$ goes on increasing until it reaches the limiting
value of 0.688.  

If we examine the phase diagram in Fig. \ref{phase}, we can see that for any fixed value of
$\phi$ less than  0.66, 
the colloidal  system continues to be in modulated liquid
state for all values of $\beta V_{o}$. For values larger than 0.7, the system remains in crystalline form
throughout the entire range of field strengths. The intermediate region is
interesting. For example, consider a $\phi$ value of 0.675. For low field strength,
at this volume fraction, the system behaves like a modulated liquid. If one
exceeds a certain field strength (
$\beta V_{o} = 1.4$ ), the system undergoes a transition to a
modulated crystalline state. But beyond $\beta V_{o} = 31$, the
hard sphere system again becomes a modulated liquid. Thus, from the measurement
of the translational order parameter in Monte Carlo simulations, we find a
possibility of re-entrant behaviour in the two-dimensional hard sphere system
in the presence of an external modulation potential.

To  examine the three phases at $\phi =
0.675$, we have measured in our simulations the equilibrium
real-space particle density $<\rho(x,y)>$ for three different field strengths,
as shown in Fig\ref{con1}(a)-(c).  After equilibrating for $10^{5}$
configurations,  $<\rho(x,y)>$ was measured by averaging over $1.5 \times
10^{5}$ configurations. From the plots, we
can see that for $\beta V_{o}$ equal to $0.1$ , the system is in a liquid
state, the modulations not being very prominent. When $\beta V_{o}$ is equal to $2.0$, 
the system is nearly crystalline and for
$ \beta V_{o} = 1000 $ , the system  loses its crystalline structure
and again becomes a modulated liquid. 
This confirms the occurrence of
a re-entrant melting as a function of the strength of the modulating potential.
This is consistent with the recent
prediction by Frey {\it et al} \cite{frey}, but it remains to be shown that the 
melting here is a continuous transition and is mediated by the unbinding of dislocations. More extensive
simulations are needed to demonstrate the latter. Meanwhile,
it would be interesting to carry out experiments involving
hard sphere colloidal particles and observe whether
such a reentrant melting occurs as the strength of the laser field is
increased. 

In conclusion, we have studied the effect of the external laser field modulation
on a two-dimensional system of colloidal particles interacting via a hard sphere potential. By identifying the maxima in the rate of variation of the order-parameter
$\rho_d$ ( as a function of the packing fraction ) as the onset of condensation
of the density mode, we have constructed the phase diagram for the system.
For a certain range of packing fractions, re-entrant melting occurs in the system
as the strength of the field increases. Measurement of real-space density for $\phi = 0.675$  confirms that the system first freezes into crystalline form but
later on becomes a modulated liquid on increase of the field strength. 
A method has been proposed recently to calculate isothermal elastic constants for the hard disk triangular solids in two dimensions using Monte Carlo simulations \cite{sen}. It would be of interest to check (using this method) whether 
these elastic constants soften when the field strength is increased beyond
a certain value in our systems.

After our work was completed, we learnt of an independent and more extensive
Monte Carlo simulational study on the same system by Strepp {\it et al}\cite{kon}. Using the the cumulant intersection
method, they have obtained a phase diagram has been obtained which is qualitatively and quantitatively similar
to the one obtained by us. This further strengthens the case for the existence of the re-entrant
modulated liquid phase in a hard-sphere colloidal system subject to an
external laser field modulation.

\section{Acknowledgements}
We are grateful to Chinmay Das for helpful discussions.

\newpage
\begin{figure}
\caption{\label{z1}Variation of $\rho_{d}$ 
 and  $d\rho_{d}/d\phi$ for $\beta V_{o}$ = 0 and for infinte field .
}
\end{figure}    
\begin{figure}
\caption{\label{t2}  Plots of $\rho_{d}$ and  $d\rho_{d}/d\phi$ 
for some typical values of field strength, $\beta V_{o}
= 0.1, 1.0, 10.0, 2000.0$ .
}
\end{figure}  
\begin{figure}
\caption{\label{phase} Phase diagram of the colloidal system for 900 particles in an external
laser field. 
}
\end{figure}  
\begin{figure}
\caption{\label{con1}Contour plots of
average density for $\phi=0.675$ in presence of external field. (a), (b)
and (c) correspond to  $\beta V_{o} = 0.1, 2.0, 1000.0$ respectively.
}
\end{figure} 
\newpage
\begin{center}
\epsfig{file=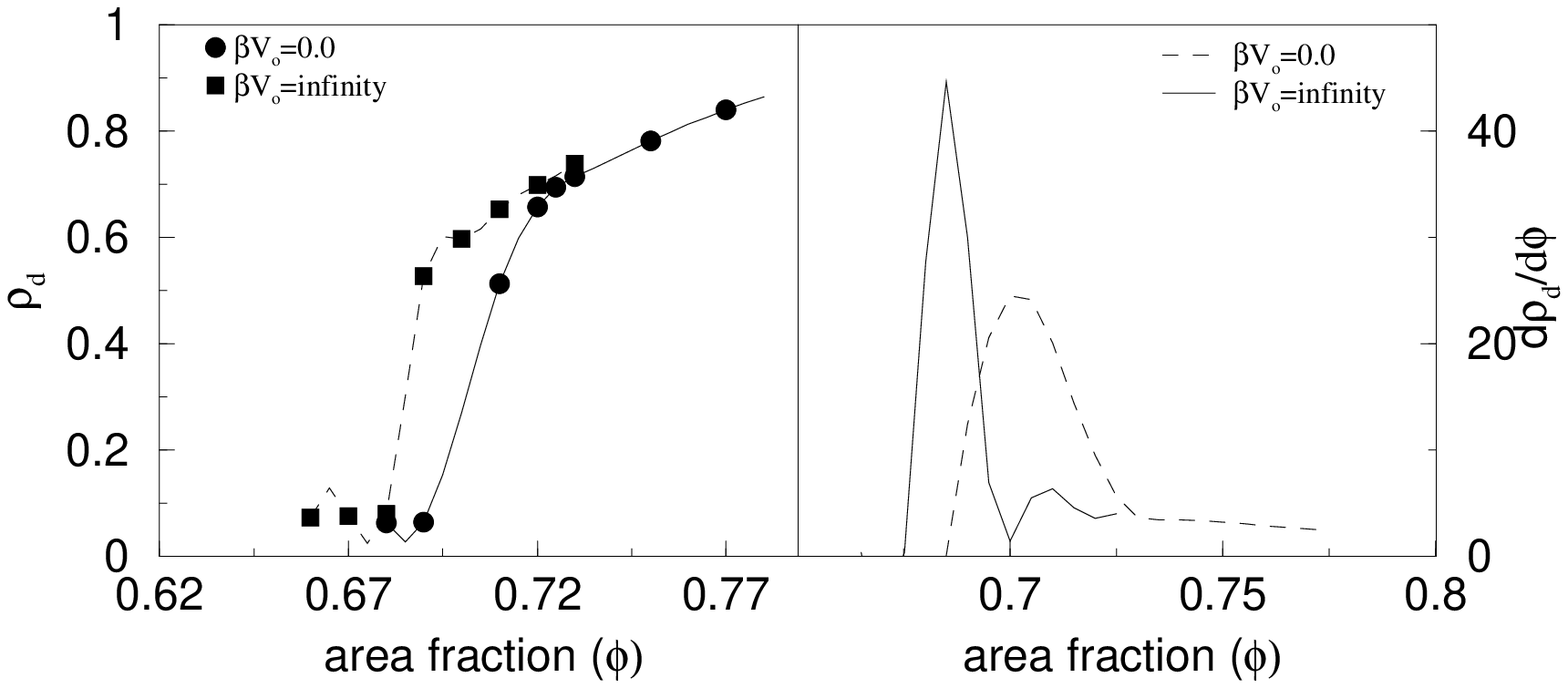}
{\bf Fig.1}
\end{center}
\newpage
\begin{center}
\epsfig{file=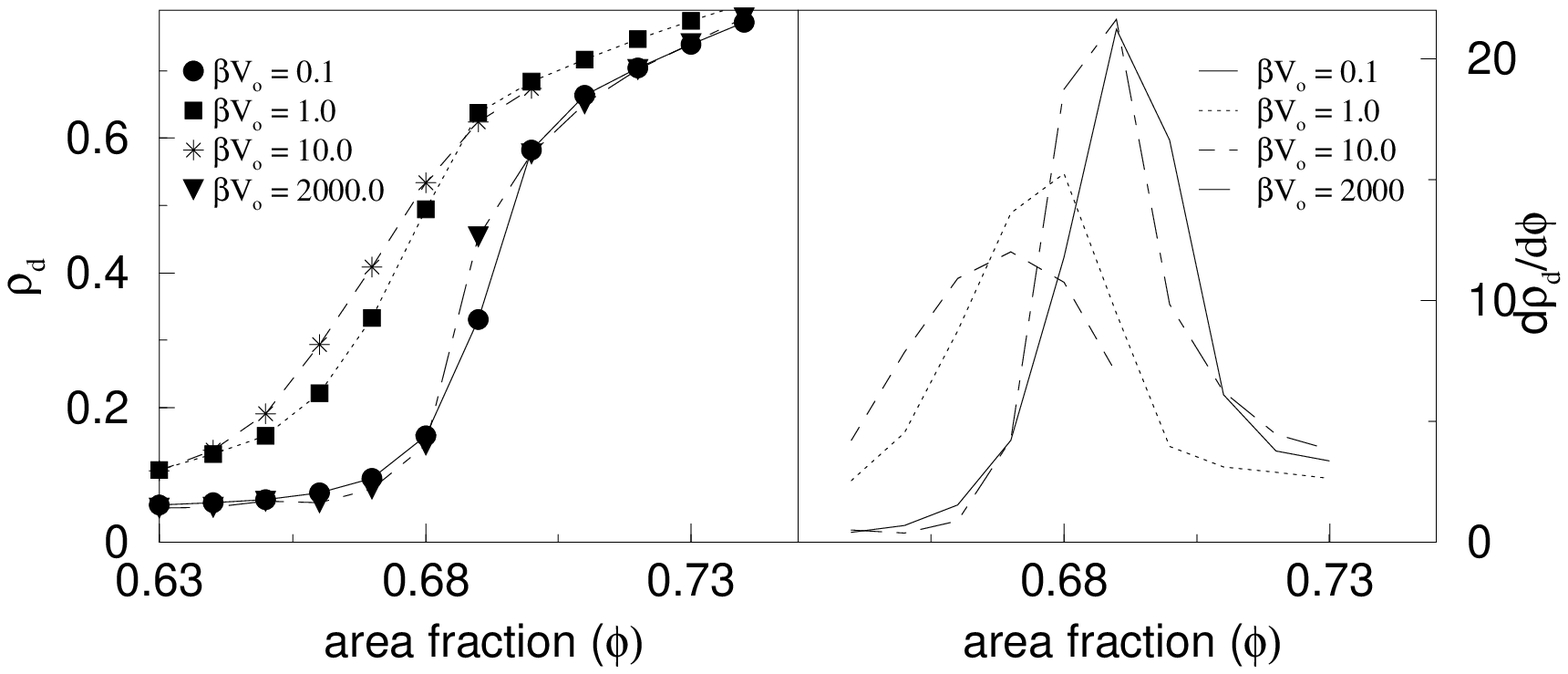}
{\bf Fig.2}
\end{center}
\newpage
\begin{center}
\epsfig{file=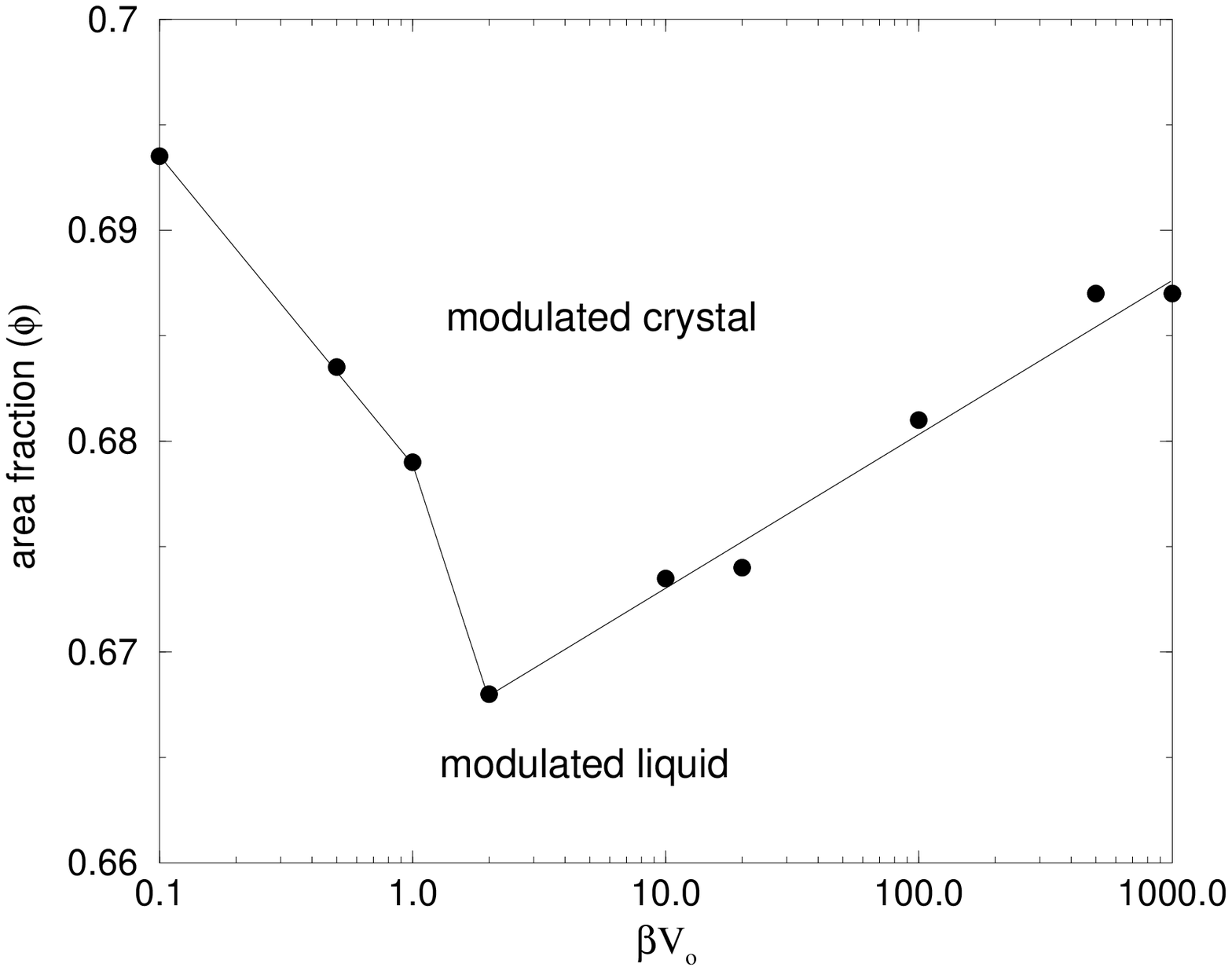,height=14cm}
{\bf Fig.3}
\end{center}
\newpage
\begin{center}
\psfig{file=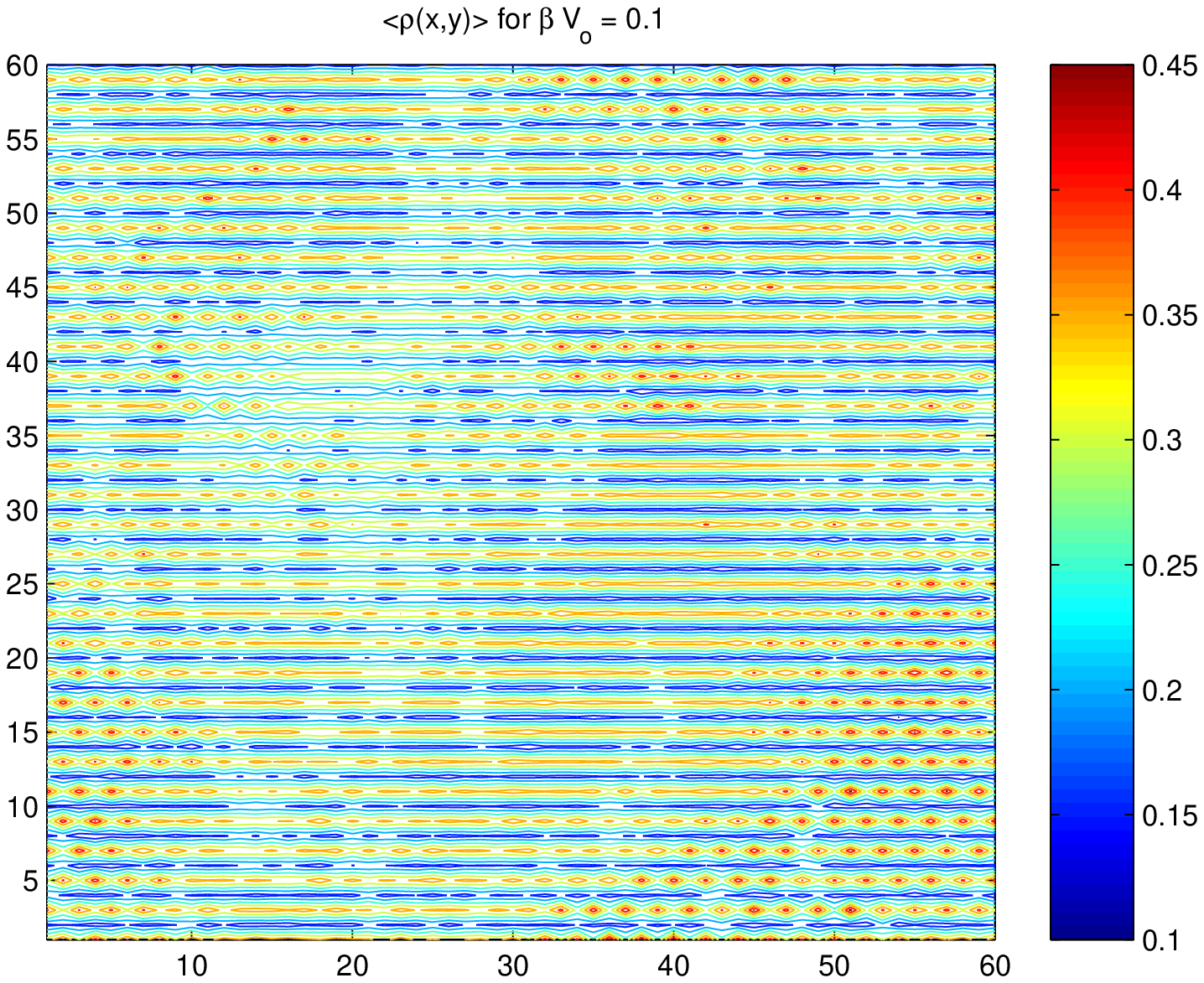,height=14cm}
{\bf Fig.4a}
\end{center}
\newpage
\begin{center}
\psfig{file=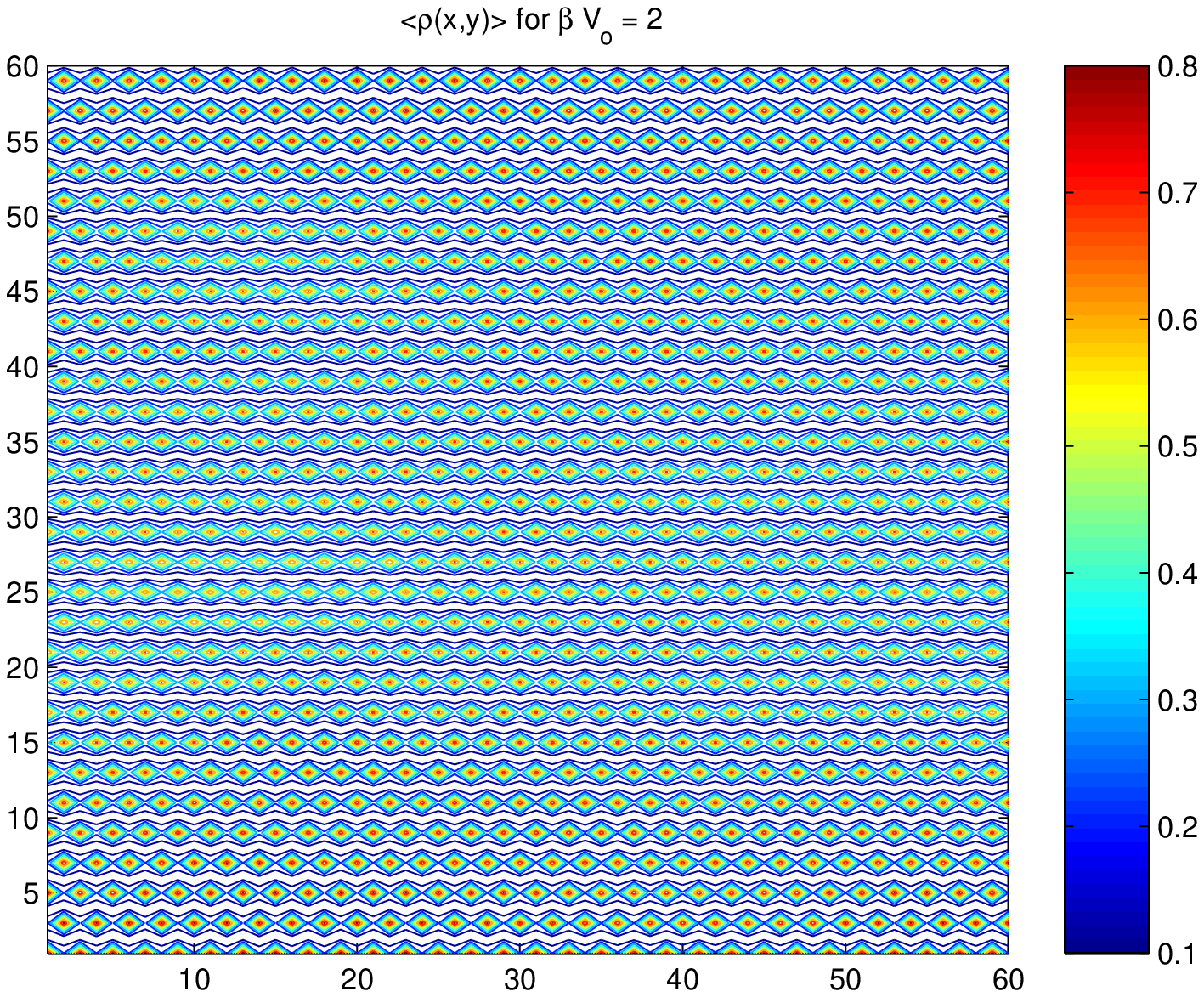,height=14cm}
{\bf Fig.4b}
\end{center}
\newpage
\begin{center}
\psfig{file=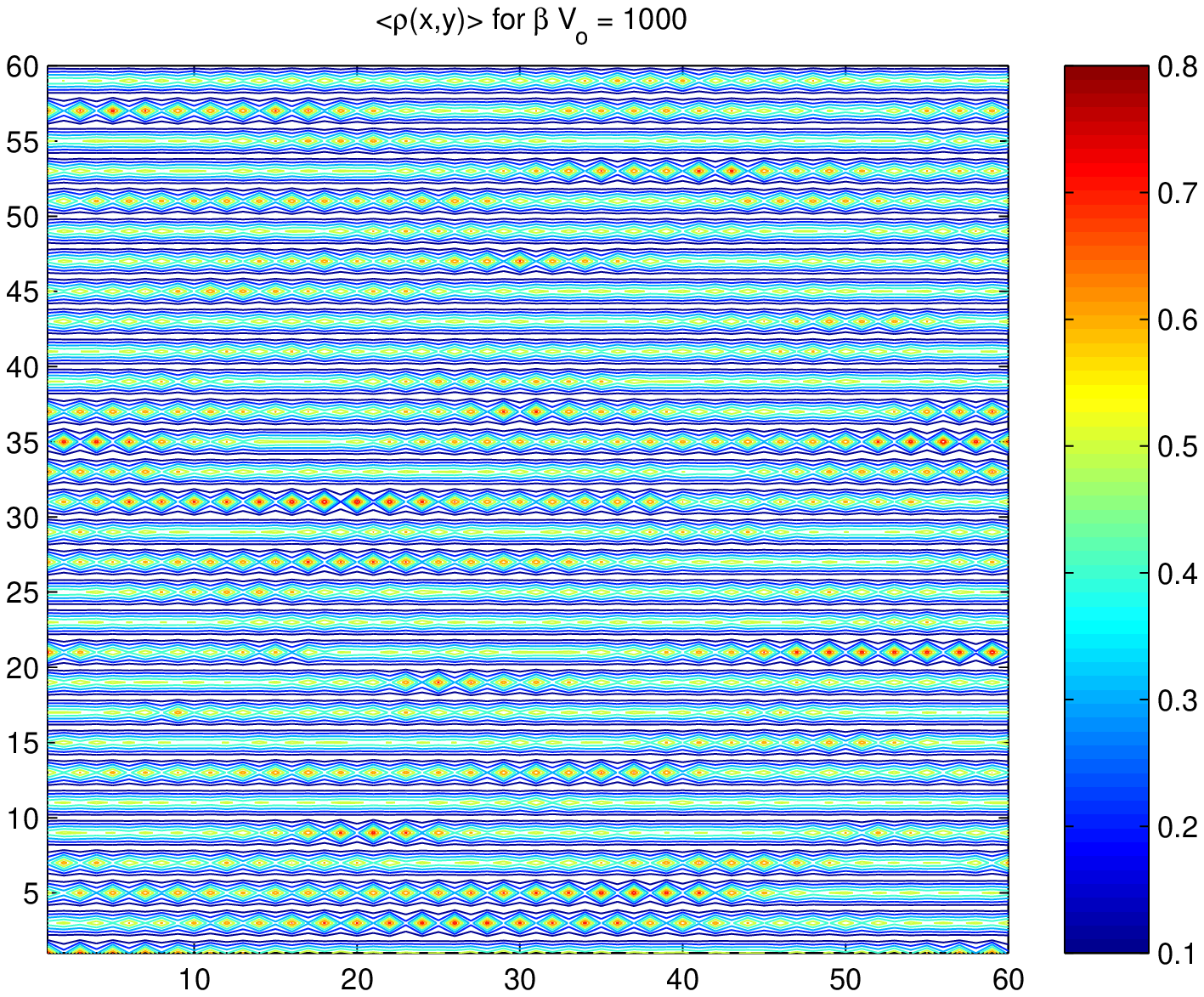,height=14cm}
{\bf Fig.4c}
\end{center}
\end{document}